\documentclass[11pt]{article}
\usepackage[margin=1in]{geometry}
\usepackage{amsmath,amssymb}
\usepackage{booktabs,tabularx,threeparttable,longtable,array}
\usepackage{graphicx}
\usepackage{natbib}
\usepackage[protrusion=true,expansion=false]{microtype}
\usepackage{setspace}
\usepackage{hyperref}
\hypersetup{hidelinks}
\usepackage{enumitem}
\usepackage{caption}
\usepackage{float}
\usepackage{pdflscape}
\title{Modelling Robust Lending Decisions under Climate Scenario Ambiguity: A Minimax-Regret Framework with NGFS Short-Term Scenarios}
\author{M. Palaisti}
\date{}

\begin{document}
	\maketitle
	
	\begin{abstract}
		This paper develops a public-data framework for evaluating incremental bank lending when plausible climate scenarios imply different sector credit outcomes but defensible scenario probabilities are unavailable. The calibration combines 2025 Shared National Credit industry commitments, January 2026 U.S. industry leverage, interest coverage and equity volatility from Damodaran, and official 2027 sector-level probability-of-default adjustments from the NGFS CLIMACRED model. One-year Merton probabilities provide independent baseline credit-risk proxies for 16 non-financial sectors representing \$5.2808 trillion of commitments. Minimax regret is compared with symmetric scenario weighting and maximin under two principal decision scores: expected loss and a credit-compensation payoff using Damodaran synthetic spreads. DAPS raises the commitment-weighted one-year PD from 0.0592\% to 0.0897\%. At a 12.5\% sector cap, minimax regret reduces maximum expected-loss regret by 29.1\% relative to symmetric scenario weighting; under the credit-compensation score the reduction is 69.1\%. Sensitivity to concentration limits, crosswalk weights, Merton inputs and the climate-PD transport shows that the value of minimax regret is conditional on scenario-specific sector rankings differing near binding portfolio constraints. The contribution is an operational decision-fragility framework for financial institutions and supervisors, not evidence about realized bank behavior or universally optimal climate lending.
	\end{abstract}
	
	
	
	\section{Introduction}
	
	Climate scenario analysis has become a central tool for assessing financial exposure to transition and physical risks. Its role is especially important because climate-related financial risk is forward-looking, nonlinear and only weakly identified by historical data. The Network for Greening the Financial System (NGFS) presents scenario analysis as an ``if--then'' exercise for exploring plausible futures rather than as a probability forecast. Its guidance also cautions that assigning probabilities to climate narratives can create a false sense of precision when uncertainty concerns policy, technology, physical damages and macrofinancial transmission simultaneously \citep{ngfs2025short,ngfsportal}. This distinction creates a decision problem that is easy to overlook. A bank may be able to estimate credit losses under several scenarios without having a defensible probability distribution with which to collapse those scenarios into one expected outcome.
	
	The first NGFS short-term scenarios, released in May 2025, sharpen this problem. They were designed for stress testing and macrofinancial risk assessment over the next several years and combine abrupt physical shocks, transition-policy changes, financial amplification and sectoral transmission in four narratives: Highway to Paris (HWTP), Sudden Wake-Up Call (SWUC), Disasters and Policy Stagnation (DAPS), and Diverging Realities (DiRe) \citep{ngfs2025short}. Subsequent NGFS work has shown that alternative macroeconomic modelling frameworks can generate different quantitative readings of the same narratives, reinforcing the relevance of model uncertainty as well as scenario uncertainty \citep{ngfs2026twolenses}. \citet{pui2026} document scenario, model and vintage uncertainty in NGFS projections more directly.
	
	This paper asks a deliberately narrower operational question: \emph{How could a bank evaluate an incremental sector-lending allocation when several plausible NGFS scenarios support different credit outcomes and the bank does not want to assign them artificial probabilities?} The object is not observed bank behavior, causal identification or a claim about the universally optimal lending portfolio. It is the fragility of a decision made before the scenario is known. If a lending allocation is optimized for one scenario, how much is lost if another scenario materializes? Can an allocation remain close to the scenario-specific optimum across the scenario set?
	
	We formulate that question using minimax regret. For each scenario, an oracle portfolio is the allocation that would be chosen if that scenario were known in advance. Regret is the difference between oracle performance and the performance of an allocation selected ex ante. Minimax regret (MMR) minimizes the largest of these opportunity losses. It differs from symmetric scenario weighting, which imposes equal decision weights on the four narratives, and from maximin, which protects the worst absolute score. MMR instead protects the worst shortfall relative to what could have been achieved after the scenario was known.
	
	The application combines three public-data layers. First, the 2025 Shared National Credit (SNC) review supplies observed U.S. large syndicated-credit commitments by industry \citep{snc2026}. We retain 16 non-financial operating-industry categories representing $\$5.2808$ trillion of commitments and interpret the decision as the allocation of a normalized \emph{incremental} lending budget, not the liquidation and reconstruction of the existing SNC stock. Second, January 2026 U.S. industry leverage, interest coverage and equity volatility from \citet{damodaran2026} are used to construct one-year Merton baseline-PD proxies and synthetic credit-spread proxies. Third, the NGFS CLIMACRED workbook supplies official 2027 sector-level baseline PDs and PD adjustments under DAPS, HWTP, SWUC and DiRe. The climate-to-credit transmission therefore comes from CLIMACRED rather than from an author-chosen common nonlinear transformation.
	
	The paper makes three contributions. First, it provides a transparent public-data calibration of a climate-scenario decision problem rather than a synthetic-obligor experiment. Second, it places an expected-loss score and a credit-compensation score on equal footing. The expected-loss specification avoids using a synthetic spread as a profitability measure; the credit-compensation specification has a formal payoff interpretation but is explicitly treated as a proxy rather than observed loan pricing. Third, the paper shows that the relevant result is conditional rather than universal: MMR is valuable when climate scenarios change the ranking of marginal lending opportunities near binding portfolio constraints. The full sector crosswalk, score rankings, cutoff distances, regret matrices and sensitivity to crosswalk weights are therefore exposed rather than left only in code.
	
	At the benchmark $12.5\%$ sector cap, MMR lowers maximum regret from $0.0967$ to $0.0686$ basis points under the expected-loss score, a $29.1\%$ reduction. Under the credit-compensation score, maximum regret falls from $0.9345$ to $0.2891$ basis points, a $69.1\%$ reduction. These percentages should not be read as general effects of robust climate lending. The magnitude depends on the score, the mapping and the portfolio constraint. The more durable result is that scenario-conditioned decisions can differ even when scenario probabilities are intentionally unavailable, and that the cost of committing to one such decision can be quantified in the same units as the decision score.
	
	\section{Related literature and contribution}
	
	The paper sits at the intersection of climate stress testing and structural credit risk, bank lending responses to climate risk, climate-scenario uncertainty, and robust portfolio choice. The contribution is not a new climate scenario or a new default model in isolation. It is an operational decision layer that asks how fragile an incremental lending allocation is when the scenario generating sector credit outcomes is itself unknown.
	
	Structural credit models link firm value to default following \citet{merton1974} and \citet{leland1994}. Climate stress-testing research embeds transition and physical shocks into financial exposures. \citet{battiston2017} show how climate-policy shocks can propagate through financial networks, while \citet{acharya2023} emphasize the difficulty of mapping scenario narratives into short-run credit outcomes. At the loan-portfolio level, \citet{nguyen2023} combine a Merton PD model with forward-looking transition pathways for major U.S. banks, and \citet{alessi2024} study how transition shocks and balance-sheet amplification can affect bank losses and capital requirements. \citet{bouveret2025} trace carbon-price shocks through macroeconomic factors into a multi-sector credit portfolio. Closest to the present short-horizon setting, \citet{cordova2026} combine granular loan data with NGFS short-term scenarios for Chile and translate sector shocks into firm-level PDs and bank losses. \citet{jung2026} similarly combine general-equilibrium estimates of climate-policy effects with historical loan-default information to construct forward-looking measures of U.S. banks' transition-risk exposures. These studies establish heterogeneous climate-credit effects but typically condition on a specified scenario.
	
	A related empirical literature shows that banks respond to climate risk through the price, maturity, quantity and composition of credit. \citet{ehlers2022} document a carbon premium in syndicated loans after the Paris Agreement. \citet{delis2024} show that banks price climate-policy exposure in lending to fossil-fuel firms, while \citet{ivanov2024} find that firms affected by carbon-pricing policy face shorter maturities, higher rates and reduced access to permanent bank financing. Physical risk also enters lending terms \citep{kempa2026}. At the bank level, \citet{bellinvia2025} report that both transition and physical exposures are associated with higher credit risk in European banks. \citet{palaisti2026pricing} develops a banking-book climate-capital pass-through approach that converts transition-risk capital charges into borrower-level loan premia. This literature motivates treating climate risk not only as a measurement problem but also as a credit-governance problem.
	
	The present paper is also related to scenario-specific climate-aware capital allocation. \citet{palaisti2025capital} develops a structural NGFS framework in which scenario shocks alter obligor PDs and constrained lending allocations. The distinction here is substantive. Scenario-specific optimization asks what the bank would choose \emph{conditional on a selected scenario}; the present paper asks how the bank could evaluate a lending decision \emph{before it knows which plausible scenario will govern the relevant credit outcomes}. The object of interest therefore shifts from the scenario-specific optimum to decision fragility across scenarios.
	
	Climate risk itself is characterized by deep uncertainty, nonlinearities and model dependence \citep{battiston2021}. The NGFS short-term scenarios are explicitly intended as plausible ``if--then'' narratives rather than probability forecasts \citep{ngfs2025short,ngfsportal}. The later ``Two Lenses'' exercise shows that different macroeconomic models can produce different quantitative readings of the same short-term narratives \citep{ngfs2026twolenses}. \citet{pui2026} document scenario, model and vintage uncertainty, while the Bank of England has likewise emphasized sensitivity to modelling choices and damage functions \citep{boe2026}. \citet{leguenedal2025} address uncertainty over the transition scenario in corporate debt valuation, allowing information about the transition path to be learned over time.
	
	Finally, robust optimization protects decisions against parameter uncertainty by optimizing over an uncertainty set rather than a single estimated vector \citep{bental1999}. Minimax regret instead evaluates a decision relative to the action that would have been optimal after the uncertain state became known. In portfolio applications, \citet{xidonas2017} construct minimax-regret portfolios over finite return scenarios, while \citet{cacador2021} compare relative-regret and absolute-robust portfolio strategies. The uncertain state here is an NGFS climate scenario and the decision object is incremental sector lending rather than traded-asset weights.
	
	The resulting gap is narrow but operationally relevant. Climate-credit studies quantify scenario-conditioned PDs and losses; lending studies show that banks change credit terms in response to climate risk; robust-portfolio studies provide decision criteria under non-probabilistic uncertainty. This paper connects those strands by evaluating the \emph{regret of the lending allocation itself}, while keeping the public-data calibration and its limitations visible.
	
	\section{Framework and public-data calibration}
	
	\subsection{Observed lending exposures and industry crosswalk}
	
	Let $j=1,\ldots,J$ index lending sectors and $s\in\mathcal S$ the four NGFS short-term scenarios,
	\[
	\mathcal S=\{\mathrm{DAPS},\mathrm{HWTP},\mathrm{SWUC},\mathrm{DiRe}\}.
	\]
	The exposure base is the 2025 SNC Program review produced jointly by the Federal Reserve, Federal Deposit Insurance Corporation and Office of the Comptroller of the Currency \citep{snc2026}. SNC covers large credits shared by multiple regulated institutions and is not the entire U.S. commercial-loan market or a confidential bank portfolio. It does, however, provide an observed large-corporate-credit environment rather than simulated exposures.
	
	The primary sample excludes banks, other financials, insurance/fiduciaries and government/education. ``Diversified Miscellaneous'' is also excluded from the primary climate mapping because it has no defensible unique NGFS counterpart. The resulting $J=16$ industries represent $\$5.2808$ trillion of 2025 commitments. The complete SNC--NGFS crosswalk, equal weights, rationale and Damodaran constituents are reported in Appendix~\ref{app:crosswalk}. Direct mappings are used where possible; broad SNC categories use transparent composites of economically related NGFS sectors. Equal within-category weights are the benchmark, not an asserted empirical exposure decomposition. Section~\ref{sec:crosswalksens} therefore perturbs these weights over the full simplex while holding a given draw fixed across scenarios.
	
	The decision variable is
	\[
	x_j=\text{share of a normalized incremental lending budget assigned to sector }j.
	\]
	The observed SNC commitments characterize the lending environment and are used to aggregate portfolio-level PD and expected-loss quantities. They are not imposed as target decision weights.
	
	\subsection{Independent baseline credit calibration}
	
	Baseline credit risk is calibrated from the January 2026 U.S. industry dataset of \citet{damodaran2026}. For each underlying Damodaran industry, let normalized total capital equal one, debt equal $D$, equity equal $E=1-D$, and observed equity volatility equal $\sigma_E$. The standard Merton system is
	\[
	E=V\Phi(d_1)-De^{-rT}\Phi(d_2),
	\qquad
	\sigma_E=\frac{V}{E}\Phi(d_1)\sigma_V,
	\]
	where
	\[
	d_1=\frac{\log(V/D)+(r+\tfrac12\sigma_V^2)T}{\sigma_V\sqrt{T}},
	\qquad d_2=d_1-\sigma_V\sqrt{T}.
	\]
	We use $T=1$ year and $r=3.47\%$, the one-year U.S. Treasury par yield on 2 January 2026. The baseline structural probability is
	\[
	PD_j^0=\Phi(-d_2).
	\]
	Because the Merton transformation is nonlinear, the system is solved separately for every underlying Damodaran industry and the resulting PDs are then aggregated to the SNC category using Damodaran firm counts.
	
	Interest coverage is aggregated using the same weights and mapped to Damodaran's January 2026 synthetic rating/default-spread table. The resulting spread $q_j$ is a public credit-compensation proxy, not an observed bank-loan spread. The baseline PDs are likewise structural market-based proxies, not regulatory through-the-cycle PDs or empirical default frequencies. Appendix~\ref{app:calibration} reports the sector-level calibration and the Damodaran constituents used in each SNC category.
	
	\subsection{Official NGFS climate-credit overlay}
	
	For climate transmission we use the 2027 CLIMACRED outputs in the NGFS/IIASA short-term-scenario workbook. CLIMACRED reports sector-level baseline-PD and PD-adjustment fields. For NGFS sector $k$ and scenario $s$ define
	\[
	p^{NGFS,0}_{ks}=\frac{\texttt{baseline\_pd}_{ks}}{100},
	\qquad
	p^{NGFS,1}_{ks}=\frac{\texttt{baseline\_pd}_{ks}+\texttt{pd\_adjustment}_{ks}}{100}.
	\]
	The benchmark transports the official change in log odds,
	\[
	\Delta_{ks}=\operatorname{logit}(p^{NGFS,1}_{ks})-
	\operatorname{logit}(p^{NGFS,0}_{ks}).
	\]
	If SNC sector $j$ maps to NGFS sectors $k\in\mathcal K_j$ with benchmark weights $w_{jk}$, then
	\[
	\bar\Delta_{js}=\sum_{k\in\mathcal K_j}w_{jk}\Delta_{ks},
	\qquad
	PD_{js}=\operatorname{logit}^{-1}\!\left[
	\operatorname{logit}(PD_j^0)+\bar\Delta_{js}
	\right].
	\]
	The benchmark uses equal $w_{jk}$ within each mapped category. As a transport robustness check we also use
	\[
	PD^{mult}_{js}=PD_j^0
	\frac{\bar p^{NGFS,1}_{js}}{\bar p^{NGFS,0}_{js}}.
	\]
	The construction deliberately separates baseline credit calibration from climate transmission: Damodaran/Merton determines the baseline risk scale, while CLIMACRED supplies the official climate-credit shock.
	
	\subsection{Two principal lending scores}
	
	The first principal score is expected loss per unit of incremental lending,
	\[
	a^{EL}_{js}=-LGD\cdot PD_{js},
	\qquad LGD=40\%.
	\]
	This specification does not require an observed or synthetic lending margin. It also has a conventional net-return interpretation under common non-credit terms. If a unit of lending earns a common promised margin $m$ and incurs common funding, operating and capital charges $c_f,c_o,c_k$, then
	\[
	a^{net}_{js}=m-c_f-c_o-c_k-LGD\cdot PD_{js}.
	\]
	Because $\sum_jx_j=1$, the common term $m-c_f-c_o-c_k$ is constant across feasible allocations. Maximizing this net-return template is therefore decision-equivalent to minimizing expected loss. Sector-specific funding costs, operating costs and capital charges would require bank-specific data and are not fabricated here.
	
	The second principal specification uses the Damodaran synthetic spread,
	\[
	a^{CC}_{js}=q_j(1-PD_{js})-LGD\cdot PD_{js}.
	\]
	Its payoff convention is explicit: the promised credit spread $q_j$ is received in survival; in default, the spread is not received and the lender incurs an LGD loss. This makes the expected payoff internally coherent. However, because $q_j$ is itself a synthetic bond-market default-spread proxy based on credit quality, it may also reflect risk premia and other compensation. We therefore call this the \emph{credit-compensation} score, not bank profitability, and do not treat its larger percentage regret reduction as the sole headline result.
	
	\subsection{Feasible set and decision rules}
	
	The feasible set is
	\[
	\mathcal X=\left\{x\in\mathbb R_+^J:\sum_jx_j=1,\;x_j\le\kappa\right\}.
	\]
	The benchmark $\kappa=12.5\%$ equals twice the equal-weight share $1/16=6.25\%$ and forces a fully allocated portfolio to use at least eight sectors. It is an illustrative intermediate concentration policy, not a regulatory threshold. The observed SNC environment itself is more concentrated: Technology, Telecom and Media represents $21.1\%$ of mapped commitments and Real Estate and Construction $10.7\%$. Thus $12.5\%$ meaningfully limits the largest observed sector without forcing near-equal weights. We examine $\kappa\in\{7.5\%,10\%,12.5\%,15\%,20\%,25\%\}$ rather than asserting that one cap is universally relevant.
	
	For either score vector $a_s$, scenario utility is $U_s(x)=a_s'x$. The scenario oracle, regret and MMR solution are
	\[
	x_s^*\in\arg\max_{x\in\mathcal X}U_s(x),
	\qquad
	R_s(x)=U_s(x_s^*)-U_s(x),
	\]
	\[
	x^{MMR}\in\arg\min_{x\in\mathcal X}\max_{s\in\mathcal S}R_s(x).
	\]
	With linear scores, MMR is the linear program
	\[
	\min_{x,R}R
	\]
	subject to
	\[
	U_s^*-a_s'x\le R\quad\forall s,
	\qquad
	\sum_jx_j=1,
	\qquad
	0\le x_j\le\kappa.
	\]
	We compare MMR with symmetric scenario weighting, which maximizes the arithmetic mean of the four scenario utilities, and maximin, which maximizes the worst absolute scenario utility. Equal scenario weights are a normative decision benchmark, not probabilities.
	
	\section{Results}
	
	\subsection{Observed exposures and model-implied scenario credit risk}
	
	The applied calibration contains 16 mapped SNC industries and $\$5.2808$ trillion of 2025 commitments. The commitment-weighted one-year Merton baseline PD is $0.0592\%$, corresponding to $\$1.250$ billion of expected loss at $40\%$ LGD. Table~\ref{tab:scenario_pd_applied} shows the model-implied scenario credit shifts. DAPS is the most adverse in aggregate, increasing the weighted PD to $0.0897\%$ and expected loss to $\$1.895$ billion. SWUC raises the aggregate PD to $0.0634\%$. HWTP and DiRe slightly reduce the weighted PD to approximately $0.0556\%$ because the mapped CLIMACRED sector set contains both adverse and favourable adjustments.
	
	\begin{table}[H]
		\centering
		\caption{Portfolio credit risk under the NGFS short-term scenarios}
		\label{tab:scenario_pd_applied}
		\small
		\begin{tabular}{@{}lrrr@{}}
			\toprule
			Scenario & Weighted 1Y PD & Change vs. baseline (bps) & Expected loss ($\$bn$) \\
			\midrule
			Baseline & 0.0592\% & 0.000 & 1.250 \\
			DAPS & 0.0897\% & 3.052 & 1.895 \\
			HWTP & 0.0556\% & -0.361 & 1.174 \\
			SWUC & 0.0634\% & 0.417 & 1.338 \\
			DiRe & 0.0556\% & -0.363 & 1.174 \\
			\bottomrule
		\end{tabular}
		\begin{minipage}{0.94\textwidth}\footnotesize
			\textit{Notes:} Baseline PDs are one-year Merton probabilities calibrated from January 2026 Damodaran industry leverage and equity volatility. Scenario PDs transport official 2027 CLIMACRED log-odds shifts to those baselines. Expected loss uses LGD $=40\%$. These are public-data model proxies, not observed default frequencies.
		\end{minipage}
	\end{table}
	
	\begin{figure}[H]
		\centering
		\includegraphics[width=0.76\textwidth]{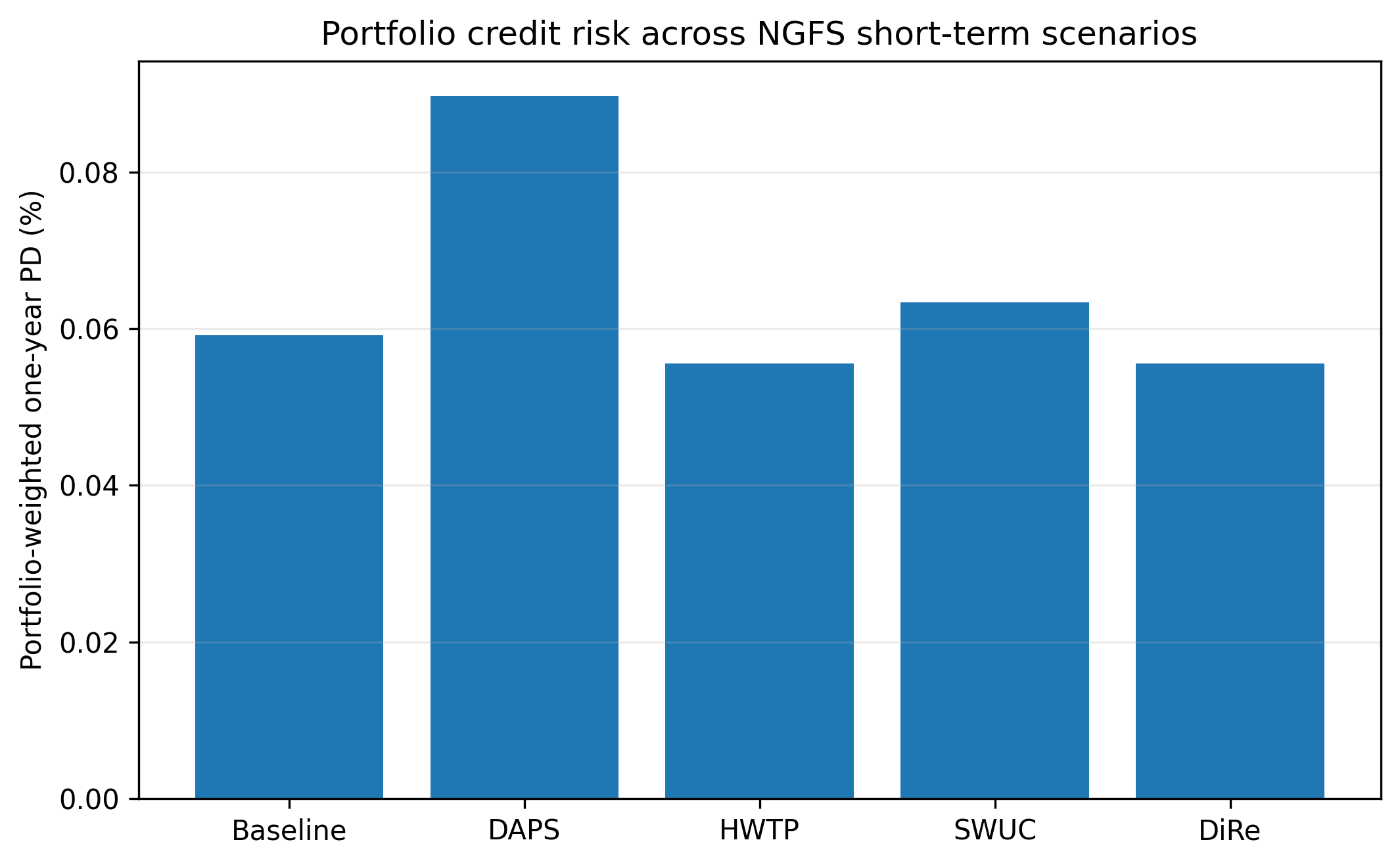}
		\caption{Commitment-weighted one-year probability of default in the baseline and four NGFS short-term scenarios.}
		\label{fig:scenario_pd}
	\end{figure}
	
	The aggregate PD ranking is not itself the decision result. What matters for allocation is whether scenario-specific sector scores change rank near the binding portfolio cutoff. The next subsections therefore present the expected-loss and credit-compensation specifications separately.
	
	\subsection{Expected-loss specification: a smaller but cleaner MMR gain}
	
	Under the expected-loss score, the average-scenario and maximin allocations have the same maximum regret at the benchmark cap, $0.0967$ basis points of expected loss. MMR lowers maximum regret to $0.0686$ basis points, a $29.1\%$ reduction. Because this score is simply $-LGD\cdot PD$, the units are directly interpretable as expected-loss basis points. The improvement is modest: on a hypothetical $\$100$ billion incremental lending envelope, the $0.0281$ basis-point reduction in worst regret is about $\$0.28$ million. The point is therefore decision stability, not a claim of large accounting savings.
	
	The scenario conflict is also transparent. The DAPS and SWUC oracles include Entertainment and Recreation among the eight capped sectors, whereas the HWTP and DiRe oracles replace that sector with Auto-Related. MMR keeps the common capped sectors but allocates $8.87\%$ to Entertainment and Recreation and $3.63\%$ to Auto-Related. Thus the expected-loss result is not driven by the synthetic spread construction.
	
	\begin{figure}[H]
		\centering
		\includegraphics[width=0.78\textwidth]{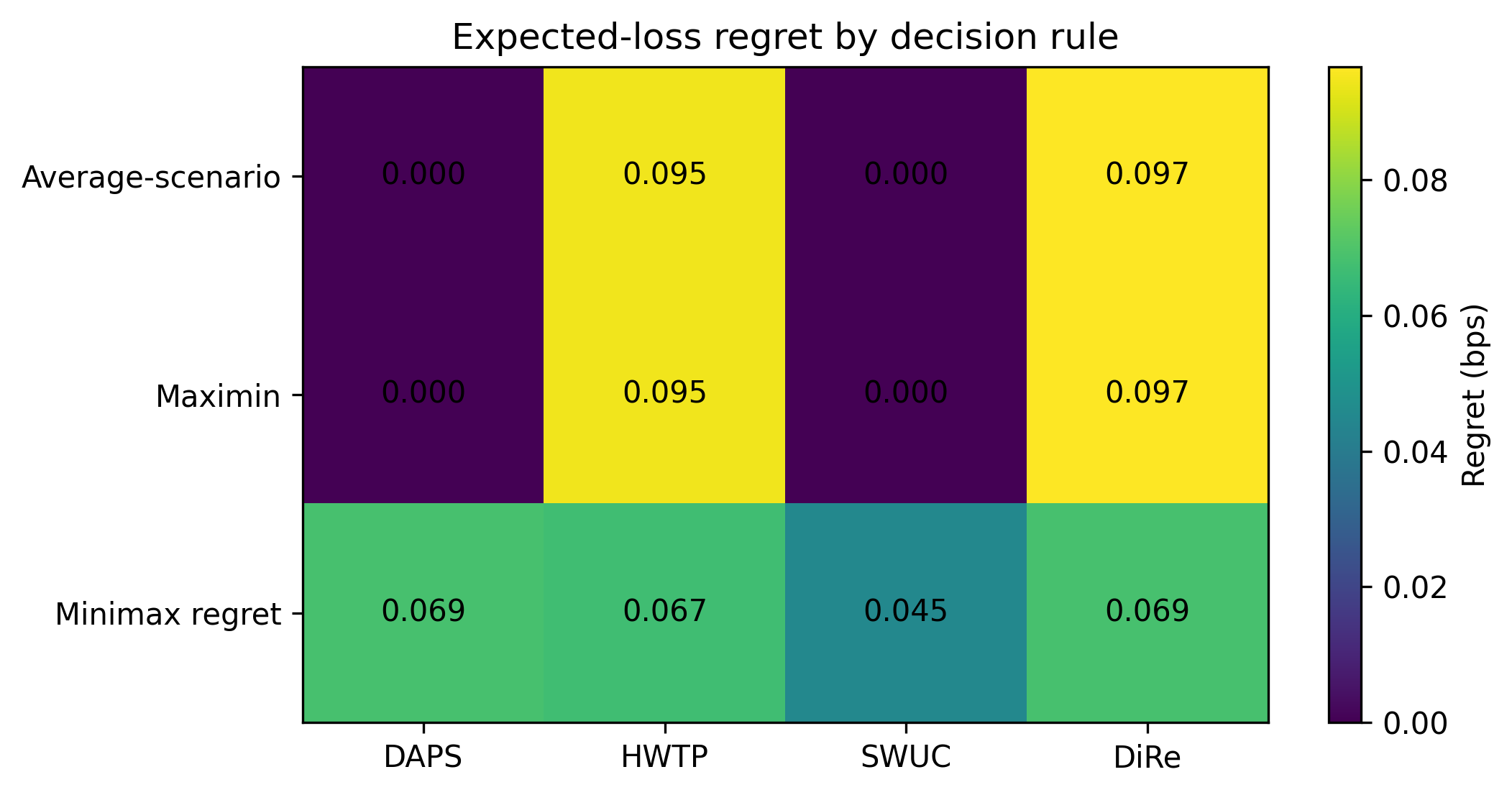}
		\caption{Scenario regret under the expected-loss specification, in basis points of expected loss.}
		\label{fig:riskregret}
	\end{figure}
	
	\subsection{Credit-compensation specification: larger regret reduction, stronger payoff dependence}
	
	Under the credit-compensation score, symmetric scenario weighting has maximum regret $0.9345$ basis points, maximin reduces it to $0.4184$, and MMR reduces it further to $0.2891$. The corresponding reductions are $69.1\%$ relative to symmetric scenario weighting and $30.9\%$ relative to maximin. Mean score declines by only $0.045$ basis points relative to the symmetric rule. The larger percentage gain is useful but should be interpreted together with the expected-loss result because the cross-sector spread proxy materially affects which sectors sit near the cutoff.
	
	\begin{table}[H]
		\centering
		\caption{Decision performance at the benchmark sector cap ($\kappa=12.5\%$)}
		\label{tab:co_primary}
		\footnotesize
		\begin{tabular}{@{}llrrrr@{}}
			\toprule
			Specification & Rule & Mean regret & Max regret & Reduction vs. avg. & Active sectors \\
			\midrule
			Expected loss & Average-scenario & 0.048 & 0.097 & -- & 8 \\
			Expected loss & Maximin & 0.048 & 0.097 & -- & 8 \\
			Expected loss & Minimax regret & 0.062 & 0.069 & 29.1\% & 9 \\
			\addlinespace
			Credit compensation & Average-scenario & 0.234 & 0.934 & -- & 8 \\
			Credit compensation & Maximin & 0.299 & 0.418 & -- & 8 \\
			Credit compensation & Minimax regret & 0.279 & 0.289 & 69.1\% & 9 \\
			\bottomrule
		\end{tabular}
		\begin{minipage}{0.94\textwidth}\footnotesize
			\textit{Notes:} Regret is in basis points of the relevant score. Expected-loss regret is directly an expected-loss quantity. Credit-compensation regret is based on $q_j(1-PD_{js})-LGD\,PD_{js}$ and should not be read as observed bank profitability. Full scenario-by-rule regret matrices, including all four scenario oracles, appear in Appendix~\ref{app:regret}.
		\end{minipage}
	\end{table}
	
	The credit-compensation mechanism is a different marginal ranking conflict. Seven sectors receive the full $12.5\%$ cap under all principal rules. DAPS ranks Food and Beverage Manufacturing and Wholesale eighth and Transportation Services tenth; HWTP, SWUC and DiRe place Transportation fourth while Food and Beverage falls to ninth. MMR therefore allocates $8.63\%$ to Food and Beverage and $3.87\%$ to Transportation rather than selecting either scenario-conditioned block completely.
	
\begin{figure}[H]
		\centering
		\includegraphics[width=0.90\textwidth]{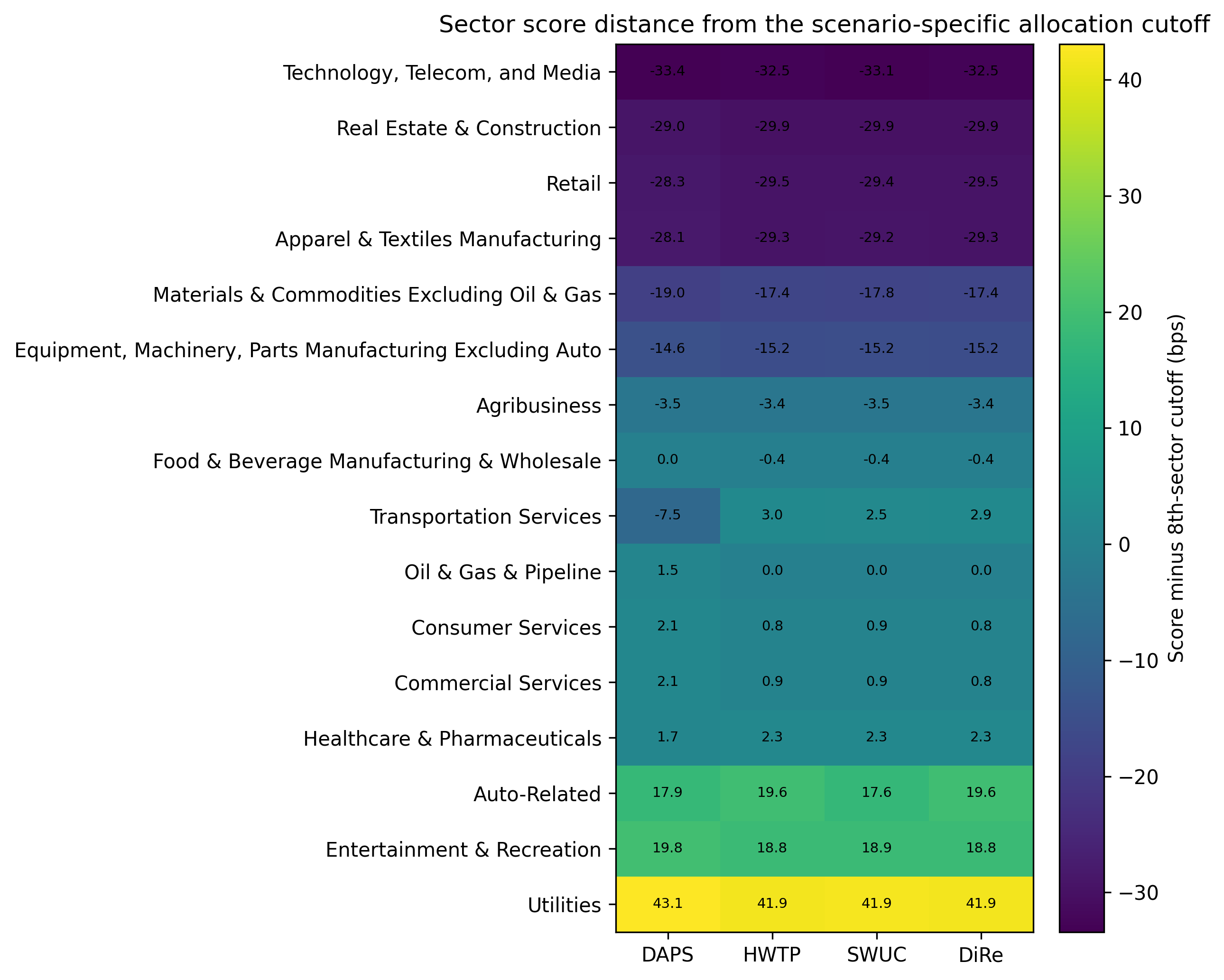}
		\caption{Credit-compensation score distance from the scenario-specific eighth-sector allocation cutoff. Positive values lie above the cutoff.}
		\label{fig:cutoff}
	\end{figure}

	Figure~\ref{fig:cutoff} reports every sector's score distance from the eighth-ranked scenario cutoff. It shows why a 16-sector linear program can still have a low-dimensional ambiguity mechanism: most sectors are far above or below the cutoff, while a small set of marginal sectors changes relative position across scenarios. This is not hidden complexity; it is the decision mechanism the regret criterion is intended to diagnose.
	
	\section{Robustness, mapping uncertainty and constraint dependence}
	
	\subsection{The value of MMR is conditional on the concentration limit}
	
\begin{figure}[H]
		\centering
		\includegraphics[width=0.78\textwidth]{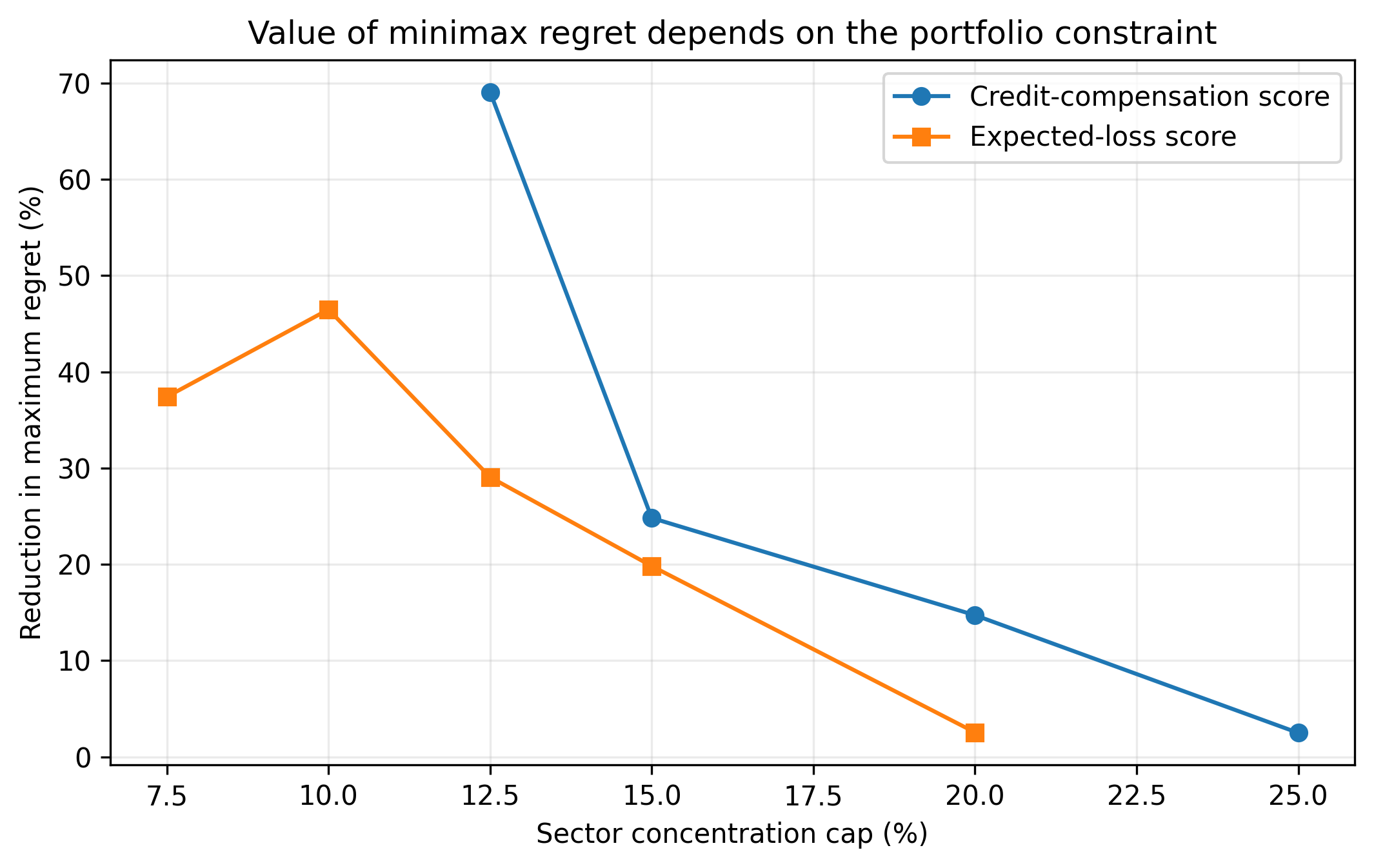}
		\caption{Reduction in maximum regret relative to symmetric scenario weighting under the two principal score specifications.}
		\label{fig:kappa_dual}
	\end{figure}

	The benchmark cap is not chosen because it maximizes the percentage gain. It is an intermediate policy that permits scenario-conditioned portfolios to differ while still imposing meaningful diversification. Figure~\ref{fig:kappa_dual} shows that the value of MMR is non-monotone in the concentration limit. Under the credit-compensation score, very tight caps of $7.5\%$ and $10\%$ essentially force the same allocation across rules, while the MMR reduction falls from $69.1\%$ at $12.5\%$ to $24.8\%$ at $15\%$, $14.7\%$ at $20\%$, and $2.5\%$ at $25\%$. Under expected loss, the corresponding reductions are $37.4\%$, $46.5\%$, $29.1\%$, $19.8\%$, $2.5\%$ and effectively zero at $25\%$.
	
	This pattern is the central conditional claim of the paper: \emph{MMR is useful when scenario-specific rankings differ near binding portfolio constraints}. If constraints are so tight that every rule is forced toward the same allocation, there is little regret to hedge. If constraints are very loose and the same high-score sectors dominate across scenarios, the ambiguity again becomes less consequential.
	
	\subsection{Crosswalk-weight uncertainty}\label{sec:crosswalksens}
	
	The equal-weight SNC--NGFS composites are a transparent benchmark but not empirically observed exposure shares. To assess whether this convenience drives the result, we draw 2,000 independent within-category weight vectors from a uniform Dirichlet distribution over each multi-sector simplex. Each draw is held fixed across the four scenarios, so the exercise changes the economic composition of an SNC category rather than choosing favourable weights scenario by scenario.
	
	The credit-compensation result is comparatively stable: the median MMR reduction in maximum regret is $69.6\%$, with a 5th--95th percentile range of $62.9\%$--$72.2\%$; the smallest reduction across the 2,000 draws is $26.1\%$. The expected-loss specification is more sensitive, as one would expect when only PD rankings determine the allocation. Its median reduction is $28.9\%$, the 5th--95th percentile range is $2.7\%$--$34.4\%$, and nearly all draws remain weakly positive apart from numerical ties. This result strengthens the conditional interpretation rather than a universal claim: when alternative credible mappings eliminate the marginal rank conflict, there is little ambiguity for MMR to resolve.
	
	\begin{figure}[H]
		\centering
		\includegraphics[width=0.78\textwidth]{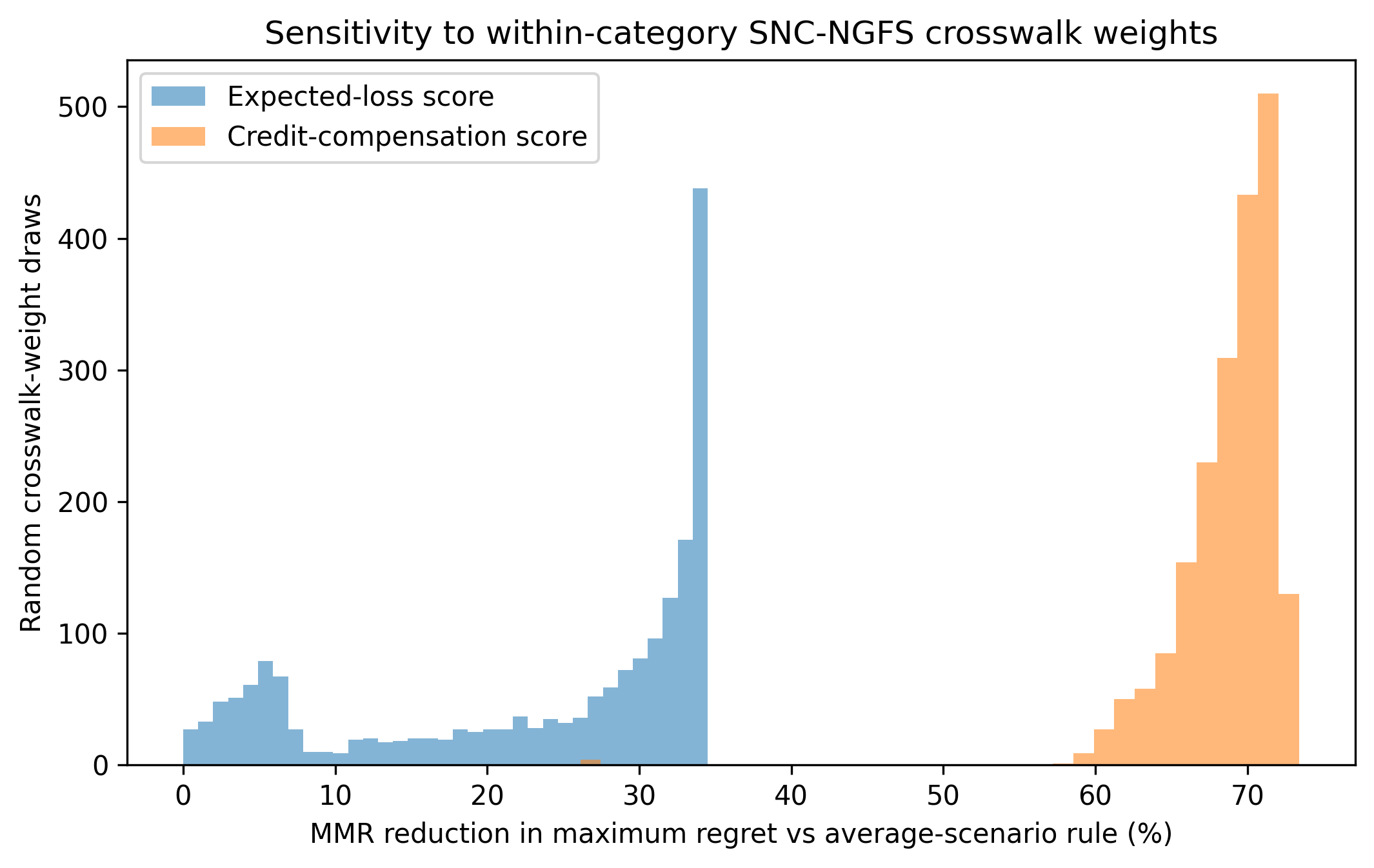}
		\caption{Sensitivity of the minimax-regret reduction to 2,000 random convex SNC--NGFS crosswalk weights.}
		\label{fig:crosswalksens}
	\end{figure}
	
	\subsection{LGD, Merton inputs and PD transport}
	
	The expected-loss percentage reduction is invariant to a common LGD because LGD scales all scenario scores proportionally: at LGD values of $20\%$, $40\%$ and $60\%$, the reduction remains $29.1\%$ while absolute regret scales linearly. Recalibrating the Merton layer at risk-free rates of $2\%$ and $5\%$ leaves the expected-loss reduction at $28.6\%$ and $29.6\%$. Scaling industry debt-to-capital ratios by $\pm10\%$ gives $29.6\%$ and $28.5\%$; scaling equity volatility by $\pm10\%$ gives $39.3\%$ and $23.5\%$. These perturbations alter baseline PD levels substantially but preserve a positive regret gain at the benchmark cap.
	
	The credit-compensation score is more payoff-sensitive. Its reduction remains between $65.8\%$ and $72.1\%$ when only the risk-free rate moves between $2\%$ and $5\%$, but some leverage, volatility and LGD perturbations make the principal scenario rankings coincide and therefore collapse regret toward zero. This is precisely why the credit-compensation specification is not allowed to carry the paper by itself.
	
	Finally, the primary log-odds climate transport is not decisive. Replacing it with the multiplicative CLIMACRED stressed-to-baseline PD ratio produces maximum regret of $0.2707$ basis points under MMR versus $0.7510$ under symmetric scenario weighting in the credit-compensation specification, a $64.0\%$ reduction. Detailed sensitivity tables are reported in Appendix~\ref{app:sensitivity}.
	
	\section{Implications for financial institutions and supervisory practice}
	
	The application supports three practical conclusions. First, climate scenario analysis can be carried beyond reporting scenario losses to examining the stability of the decisions implied by those losses. A scenario may matter operationally because it changes the ordering of marginal lending opportunities even when its portfolio-average PD effect is not the only statistic of interest. The full rank and cutoff tables in Appendix~\ref{app:regret} make this distinction observable.
	
	Second, scenario probabilities are not required for a disciplined comparison of decisions. Symmetric scenario weighting remains a useful benchmark, but equal weights are a management choice about compensability across scenarios, not probabilities. Maximin answers a different question by protecting the worst absolute score. MMR protects the worst opportunity loss relative to the scenario-specific decision that would have been chosen with hindsight. These criteria should therefore be understood as alternative governance rules.
	
	Third, the incremental-lending interpretation makes the framework compatible with ordinary risk-appetite and origination processes. Institution-specific borrower, sector, geography, rating, collateral, maturity or capital constraints can be added to the linear program. The public-data calibration does not establish what any particular bank should lend, and it does not estimate realized optimal behavior. It shows how a bank or supervisor \emph{could evaluate decision fragility} once scenario-conditioned credit scores and portfolio limits are available.
	
	These conclusions are consistent with supervisory emphasis on managing material climate-related financial risks while recognizing substantial model uncertainty \citep{bcbs2022,boe2026}. The framework is not a regulatory-capital model. It is a transparent decision and model-risk overlay on scenario-conditioned credit analysis.
	
	\section{Limitations and extensions}
	
	The study is an applied calibration, not causal or predictive bank-lending empirics. SNC supplies aggregate shared-credit commitments rather than bank-specific origination decisions or a facility-level pipeline. We do not observe contractual loan spreads, borrower default histories, facility maturity, collateral, seniority, covenants, utilization, bank-specific risk appetite, funding costs or operating costs. Accordingly, the normalized incremental budget should not be interpreted as a reconstructed SNC portfolio or a forecast of actual bank allocations.
	
	The credit inputs are also proxies. Damodaran synthetic spreads are not observed loan margins, which is why the expected-loss specification is co-primary and the credit-compensation payoff is explicitly labelled as such. Merton PDs are structural market-based proxies rather than empirical default frequencies or regulatory PDs. The weak cross-sectional association with Damodaran synthetic spreads and SNC Classified shares is therefore treated as a warning against overinterpretation, not as hidden validation.
	
	The SNC--NGFS crosswalk remains a modelling choice. The paper now reports the complete mapping, equal weights and rationale in Appendix~\ref{app:crosswalk} and systematically varies the weights over the feasible simplex. This does not substitute for borrower-level economic exposure data. Revenue, employment, output or input-output weights would be valuable where they can be constructed consistently. Alternative NGFS models, vintages and the Two Lenses outputs could also enlarge the ambiguity set beyond the four current score vectors.
	
	Finally, the optimization treats the incremental budget as divisible and abstracts from transaction costs, minimum ticket sizes, relationship constraints and endogenous repricing. These are natural extensions. Distributionally robust optimization, weighted regret, downside criteria and management-supplied scenario weights are also useful comparators for future work. The present paper deliberately keeps the primary question finite and auditable: given four official short-term scenario credit vectors and a transparent portfolio constraint, how fragile is the allocation chosen before the scenario is known?
	
	\section{Conclusion}
	
	This paper develops a public-data operational framework for evaluating incremental lending decisions under climate-scenario ambiguity. It combines observed 2025 SNC industry commitments, January 2026 U.S. industry credit fundamentals and official 2027 CLIMACRED PD adjustments for DAPS, HWTP, SWUC and DiRe. The exercise is not evidence about realized bank behavior and does not claim to identify a universally optimal climate-lending portfolio.
	
	The model-implied portfolio credit effects differ across the scenarios: DAPS raises the commitment-weighted one-year PD from $0.0592\%$ to $0.0897\%$, SWUC produces a smaller increase, and HWTP and DiRe slightly reduce it. The decision layer adds a different insight. Under the expected-loss score, MMR lowers maximum regret by $29.1\%$ at the benchmark $12.5\%$ sector cap and splits an Entertainment-versus-Auto marginal ranking conflict. Under the credit-compensation score, the reduction is $69.1\%$ and the marginal conflict is Food and Beverage versus Transportation. In both cases, MMR is doing the same economic job: it avoids committing fully to one scenario-conditioned ranking when the disputed sectors sit near a binding allocation cutoff.
	
	The size of the gain is not universal. It varies with concentration limits, crosswalk weights and the payoff specification. That dependence is part of the result. Climate-scenario ambiguity is decision-relevant when plausible scenarios change marginal rankings within the feasible region; it is much less important when constraints force common allocations or when one ranking dominates every scenario. For climate-risk governance, minimax regret therefore offers a transparent diagnostic of decision fragility rather than a claim of scenario-proof optimality.

	\appendix
	\section*{Appendices}
	The appendices make the mapping and decision mechanism auditable within the manuscript. They report the complete SNC--NGFS crosswalk and rationale, the Damodaran constituents and sector credit calibration, the full regret matrices and ranking/cutoff diagnostics, and the main sensitivity tables.
	
	\section{SNC--NGFS crosswalk and mapping rationale}\label{app:crosswalk}
	
	\begin{landscape}
		\scriptsize
		\begin{longtable}{p{3.1cm}r p{6.2cm} p{2.5cm} p{7.2cm}}
			\caption{Complete SNC--NGFS crosswalk used in the benchmark calibration}\label{tab:app_crosswalk}\\
			\toprule
			SNC industry & Commit. ($\$bn$) & Mapped NGFS sector(s) & Benchmark weights & Mapping rationale \\
			\midrule
			\endfirsthead
			\toprule
			SNC industry & Commit. ($\$bn$) & Mapped NGFS sector(s) & Benchmark weights & Mapping rationale \\
			\midrule
			\endhead
			Consumer Services & 39.1 & Market Services; Consumer Goods Industries & Equal: 0.500, 0.500 & Broad consumer-facing services; Market Services is the main service proxy and Consumer Goods captures consumer-facing operating industries. \\
			Technology, Telecom, and Media & 1113.8 & Computer, electronic and optical products; Market Services; R\&D; Paper products, publishing & Equal: 0.250, 0.250, 0.250, 0.250 & Composite covers electronics/hardware, service-heavy technology and telecom, R\&D, and publishing/media. \\
			Commercial Services & 339.8 & Market Services & Equal: 1.000 & Direct broad service-sector match. \\
			Materials \& Commodities Excluding Oil \& Gas & 439.8 & Chemical Products; Non-metallic minerals; Non-ferrous metals; Ferrous metals; Fabricated Metal products; Paper products, publishing; Rubber and plastic products & Equal: 0.143, 0.143, 0.143, 0.143, 0.143, 0.143, 0.143 & Composite across principal materials-processing and intermediate-goods sectors. \\
			Entertainment \& Recreation & 183.2 & Market Services & Equal: 1.000 & No dedicated NGFS recreation category; Market Services is the closest service proxy. \\
			Apparel \& Textiles Manufacturing & 41.8 & Consumer Goods Industries & Equal: 1.000 & Consumer Goods Industries is the closest NGFS manufacturing category. \\
			Healthcare \& Pharmaceuticals & 350.6 & Basic pharmaceutical products; Market Services & Equal: 0.500, 0.500 & Two-part composite separates pharmaceutical manufacturing from healthcare services. \\
			Food \& Beverage Manufacturing \& Wholesale & 241.9 & Consumer Goods Industries & Equal: 1.000 & Consumer Goods Industries is the closest downstream manufacturing/wholesale category; Agribusiness is separate. \\
			Equipment, Machinery, Parts Manufacturing Excluding Auto & 395.6 & Other Equipment Goods; Fabricated Metal products; Advanced Electric Appliances & Equal: 0.333, 0.333, 0.333 & Composite captures general machinery, fabricated components and electrical equipment while excluding autos. \\
			Real Estate \& Construction & 566.1 & Construction & Equal: 1.000 & Construction is the closest direct NGFS activity; Damodaran constituents retain the broader real-estate component of SNC. \\
			Transportation Services & 144.0 & Land transport; Water transport; Air transport; Warehousing & Equal: 0.250, 0.250, 0.250, 0.250 & Direct modal decomposition into land, water, air and warehousing. \\
			Agribusiness & 40.2 & Agriculture & Equal: 1.000 & Direct Agriculture match. \\
			Retail & 349.3 & Market Services & Equal: 1.000 & Retail is treated as a market-service activity rather than manufacturing. \\
			Auto-Related & 252.3 & EV Transport Equipment; Transport equipment (excluding EV) & Equal: 0.500, 0.500 & Retains both EV and non-EV transport-equipment channels. \\
			Utilities & 475.5 & Power Supply & Equal: 1.000 & Power Supply is the broad NGFS utility match. \\
			Oil \& Gas \& Pipeline & 307.8 & Crude Oil; Oil; Gas & Equal: 0.333, 0.333, 0.333 & Composite spans crude oil, oil activities and natural gas. \\
		\end{longtable}
	\end{landscape}
	
	\section{Baseline calibration and final scenario PDs}\label{app:calibration}
	
	\begin{landscape}
		\scriptsize
		\begin{longtable}{p{4.3cm}rrrrrr}
			\caption{Sector-level credit calibration}\label{tab:app_calibration}\\
			\toprule
			SNC industry & Baseline PD (\%) & Spread (bps) & DAPS PD (\%) & HWTP PD (\%) & SWUC PD (\%) & DiRe PD (\%) \\
			\midrule
			\endfirsthead
			\toprule
			SNC industry & Baseline PD (\%) & Spread (bps) & DAPS PD (\%) & HWTP PD (\%) & SWUC PD (\%) & DiRe PD (\%) \\
			\midrule
			\endhead
			Consumer Services & 0.0001 & 70.0 & 0.0001 & 0.0001 & 0.0001 & 0.0001 \\
			Technology, Telecom, and Media & 0.0963 & 40.0 & 0.1358 & 0.0835 & 0.1003 & 0.0835 \\
			Commercial Services & 0.0000 & 70.0 & 0.0000 & 0.0000 & 0.0000 & 0.0000 \\
			Materials \& Commodities Excluding Oil \& Gas & 0.0854 & 55.0 & 0.1510 & 0.0813 & 0.0910 & 0.0814 \\
			Entertainment \& Recreation & 0.0248 & 89.0 & 0.0314 & 0.0261 & 0.0265 & 0.0261 \\
			Apparel \& Textiles Manufacturing & 0.0032 & 40.0 & 0.0056 & 0.0033 & 0.0035 & 0.0033 \\
			Healthcare \& Pharmaceuticals & 0.1551 & 78.0 & 0.2057 & 0.1602 & 0.1623 & 0.1602 \\
			Food \& Beverage Manufacturing \& Wholesale & 0.0292 & 70.0 & 0.0510 & 0.0307 & 0.0319 & 0.0307 \\
			Equipment, Machinery, Parts Manufacturing Excluding Auto & 0.0267 & 55.0 & 0.0408 & 0.0258 & 0.0268 & 0.0258 \\
			Real Estate \& Construction & 0.0214 & 40.0 & 0.0270 & 0.0195 & 0.0215 & 0.0195 \\
			Transportation Services & 0.3958 & 89.0 & 0.6983 & 0.4133 & 0.4264 & 0.4133 \\
			Agribusiness & 0.1060 & 70.0 & 0.1372 & 0.1043 & 0.1082 & 0.1043 \\
			Retail & 0.0081 & 40.0 & 0.0103 & 0.0085 & 0.0087 & 0.0085 \\
			Auto-Related & 0.0479 & 89.0 & 0.0786 & 0.0071 & 0.0576 & 0.0067 \\
			Utilities & 0.0000 & 111.0 & 0.0000 & 0.0000 & 0.0000 & 0.0000 \\
			Oil \& Gas \& Pipeline & 0.0090 & 70.0 & 0.0151 & 0.0209 & 0.0230 & 0.0209 \\
		\end{longtable}
	\end{landscape}
	
	\begin{landscape}
		\scriptsize
		\begin{longtable}{p{5cm}p{18cm}}
			\caption{Damodaran industries included in each SNC calibration bucket}\label{tab:app_damodaran}\\
			\toprule
			SNC industry & Damodaran constituent industries \\
			\midrule
			\endfirsthead
			\toprule
			SNC industry & Damodaran constituent industries \\
			\midrule
			\endhead
			Consumer Services & Business \& Consumer Services; Restaurant/Dining \\
			Technology, Telecom, and Media & Computer Services; Computers/Peripherals; Semiconductor; Software (Internet); Software (System \& Application); Telecom (Wireless); Telecom. Equipment; Telecom. Services; Broadcasting; Publishing \& Newspapers; Information Services \\
			Commercial Services & Business \& Consumer Services \\
			Materials \& Commodities Excluding Oil \& Gas & Chemical (Basic); Chemical (Diversified); Chemical (Specialty); Building Materials; Metals \& Mining; Steel; Packaging \& Container; Paper/Forest Products; Rubber\& Tires \\
			Entertainment \& Recreation & Entertainment; Recreation; Hotel/Gaming \\
			Apparel \& Textiles Manufacturing & Apparel; Shoe \\
			Healthcare \& Pharmaceuticals & Drugs (Biotechnology); Drugs (Pharmaceutical); Healthcare Products; Healthcare Support Services; Hospitals/Healthcare Facilities \\
			Food \& Beverage Manufacturing \& Wholesale & Food Processing; Food Wholesalers; Beverage (Alcoholic); Beverage (Soft) \\
			Equipment, Machinery, Parts Manufacturing Excluding Auto & Machinery; Electrical Equipment; Electronics (General); Office Equipment \& Services \\
			Real Estate \& Construction & Engineering/Construction; Construction Supplies; Homebuilding; Real Estate (Development); Real Estate (General/Diversified); Real Estate (Operations \& Services); R.E.I.T. \\
			Transportation Services & Air Transport; Transportation; Transportation (Railroads); Trucking \\
			Agribusiness & Farming/Agriculture \\
			Retail & Retail (Building Supply); Retail (Distributors); Retail (General); Retail (Grocery and Food); Retail (Special Lines) \\
			Auto-Related & Auto \& Truck; Auto Parts \\
			Utilities & Power; Utility (General); Utility (Water) \\
			Oil \& Gas \& Pipeline & Oil/Gas (Integrated); Oil/Gas (Production and Exploration); Oil/Gas Distribution; Oilfield Svcs/Equip. \\
		\end{longtable}
	\end{landscape}
	
	\section{Full regret matrices and sector rankings}\label{app:regret}
	
	\begin{table}[H]
		\centering
		\caption{Full expected-loss regret matrix at $\kappa=12.5\%$ (bps)}
		\label{tab:app_risk_regret}
		\scriptsize
		\begin{tabular}{@{}lrrrrr@{}}
			\toprule
			Portfolio & DAPS & HWTP & SWUC & DiRe & Maximum \\
			\midrule
			Average-scenario & 0.000 & 0.095 & 0.000 & 0.097 & 0.097 \\
			Maximin & 0.000 & 0.095 & 0.000 & 0.097 & 0.097 \\
			Minimax regret & 0.069 & 0.067 & 0.045 & 0.069 & 0.069 \\
			Oracle: DAPS & 0.000 & 0.095 & 0.000 & 0.097 & 0.097 \\
			Oracle: HWTP & 0.236 & 0.000 & 0.155 & 0.000 & 0.236 \\
			Oracle: SWUC & 0.000 & 0.095 & 0.000 & 0.097 & 0.097 \\
			Oracle: DiRe & 0.236 & 0.000 & 0.155 & 0.000 & 0.236 \\
			\bottomrule
		\end{tabular}
	\end{table}
	
	\begin{table}[H]
		\centering
		\caption{Full credit-compensation regret matrix at $\kappa=12.5\%$ (bps)}
		\label{tab:app_credit_regret}
		\scriptsize
		\begin{tabular}{@{}lrrrrr@{}}
			\toprule
			Portfolio & DAPS & HWTP & SWUC & DiRe & Maximum \\
			\midrule
			Average-scenario & 0.934 & 0.000 & -0.000 & 0.000 & 0.934 \\
			Maximin & 0.000 & 0.418 & 0.358 & 0.418 & 0.418 \\
			Minimax regret & 0.289 & 0.289 & 0.247 & 0.289 & 0.289 \\
			Oracle: DAPS & 0.000 & 0.418 & 0.358 & 0.418 & 0.418 \\
			Oracle: HWTP & 0.934 & 0.000 & -0.000 & 0.000 & 0.934 \\
			Oracle: SWUC & 0.934 & 0.000 & -0.000 & 0.000 & 0.934 \\
			Oracle: DiRe & 0.934 & 0.000 & -0.000 & 0.000 & 0.934 \\
			\bottomrule
		\end{tabular}
	\end{table}
	
	\begin{landscape}
		\scriptsize
		\begin{longtable}{p{4.6cm}rr rr rr rr}
			\caption{Credit-compensation sector ranks and distance from the scenario cutoff}\label{tab:app_credit_ranks}\\
			\toprule
			& \multicolumn{2}{c}{DAPS} & \multicolumn{2}{c}{HWTP} & \multicolumn{2}{c}{SWUC} & \multicolumn{2}{c}{DiRe} \\
			SNC industry & Rank & Dist. & Rank & Dist. & Rank & Dist. & Rank & Dist. \\
			\midrule
			\endfirsthead
			\toprule
			& \multicolumn{2}{c}{DAPS} & \multicolumn{2}{c}{HWTP} & \multicolumn{2}{c}{SWUC} & \multicolumn{2}{c}{DiRe} \\
			SNC industry & Rank & Dist. & Rank & Dist. & Rank & Dist. & Rank & Dist. \\
			\midrule
			\endhead
			Consumer Services & 5 & 2.07 & 7 & 0.85 & 7 & 0.93 & 7 & 0.85 \\
			Technology, Telecom, and Media & 16 & -33.41 & 16 & -32.52 & 16 & -33.12 & 16 & -32.52 \\
			Commercial Services & 4 & 2.08 & 6 & 0.85 & 6 & 0.94 & 6 & 0.85 \\
			Materials \& Commodities Excluding Oil \& Gas & 12 & -19.04 & 12 & -17.45 & 12 & -17.75 & 12 & -17.45 \\
			Entertainment \& Recreation & 2 & 19.79 & 3 & 18.79 & 2 & 18.85 & 3 & 18.78 \\
			Apparel \& Textiles Manufacturing & 13 & -28.15 & 13 & -29.28 & 13 & -29.20 & 13 & -29.28 \\
			Healthcare \& Pharmaceuticals & 6 & 1.69 & 5 & 2.32 & 5 & 2.32 & 5 & 2.32 \\
			Food \& Beverage Manufacturing \& Wholesale & 8 & 0.00 & 9 & -0.40 & 9 & -0.36 & 9 & -0.40 \\
			Equipment, Machinery, Parts Manufacturing Excluding Auto & 11 & -14.58 & 11 & -15.19 & 11 & -15.15 & 11 & -15.19 \\
			Real Estate \& Construction & 15 & -29.01 & 15 & -29.94 & 15 & -29.93 & 15 & -29.94 \\
			Transportation Services & 10 & -7.48 & 4 & 2.95 & 4 & 2.50 & 4 & 2.95 \\
			Agribusiness & 9 & -3.51 & 10 & -3.40 & 10 & -3.47 & 10 & -3.40 \\
			Retail & 14 & -28.34 & 14 & -29.49 & 14 & -29.41 & 14 & -29.49 \\
			Auto-Related & 3 & 17.86 & 2 & 19.56 & 3 & 17.58 & 2 & 19.58 \\
			Utilities & 1 & 43.08 & 1 & 41.85 & 1 & 41.94 & 1 & 41.85 \\
			Oil \& Gas \& Pipeline & 7 & 1.46 & 8 & 0.00 & 8 & 0.00 & 8 & 0.00 \\
		\end{longtable}
		\par\footnotesize Distances are score minus the eighth-ranked scenario cutoff, in basis points of the credit-compensation score.
	\end{landscape}
	
	\begin{landscape}
		\scriptsize
		\begin{longtable}{p{4.6cm}rr rr rr rr}
			\caption{Expected-loss sector ranks and distance from the scenario cutoff}\label{tab:app_risk_ranks}\\
			\toprule
			& \multicolumn{2}{c}{DAPS} & \multicolumn{2}{c}{HWTP} & \multicolumn{2}{c}{SWUC} & \multicolumn{2}{c}{DiRe} \\
			SNC industry & Rank & Dist. & Rank & Dist. & Rank & Dist. & Rank & Dist. \\
			\midrule
			\endfirsthead
			\toprule
			& \multicolumn{2}{c}{DAPS} & \multicolumn{2}{c}{HWTP} & \multicolumn{2}{c}{SWUC} & \multicolumn{2}{c}{DiRe} \\
			SNC industry & Rank & Dist. & Rank & Dist. & Rank & Dist. & Rank & Dist. \\
			\midrule
			\endhead
			Consumer Services & 3 & 1.25 & 3 & 0.83 & 3 & 1.06 & 3 & 0.83 \\
			Technology, Telecom, and Media & 12 & -4.18 & 13 & -2.51 & 13 & -2.95 & 13 & -2.51 \\
			Commercial Services & 2 & 1.26 & 2 & 0.84 & 2 & 1.06 & 2 & 0.84 \\
			Materials \& Commodities Excluding Oil \& Gas & 14 & -4.78 & 12 & -2.42 & 12 & -2.58 & 12 & -2.42 \\
			Entertainment \& Recreation & 8 & 0.00 & 10 & -0.21 & 8 & 0.00 & 10 & -0.21 \\
			Apparel \& Textiles Manufacturing & 4 & 1.03 & 4 & 0.70 & 4 & 0.92 & 4 & 0.70 \\
			Healthcare \& Pharmaceuticals & 15 & -6.97 & 15 & -5.57 & 15 & -5.43 & 15 & -5.57 \\
			Food \& Beverage Manufacturing \& Wholesale & 10 & -0.79 & 11 & -0.39 & 10 & -0.22 & 11 & -0.39 \\
			Equipment, Machinery, Parts Manufacturing Excluding Auto & 9 & -0.38 & 9 & -0.19 & 9 & -0.01 & 9 & -0.19 \\
			Real Estate \& Construction & 7 & 0.18 & 7 & 0.06 & 6 & 0.20 & 7 & 0.06 \\
			Transportation Services & 16 & -26.68 & 16 & -15.70 & 16 & -16.00 & 16 & -15.70 \\
			Agribusiness & 13 & -4.23 & 14 & -3.34 & 14 & -3.27 & 14 & -3.34 \\
			Retail & 5 & 0.85 & 6 & 0.50 & 5 & 0.71 & 6 & 0.50 \\
			Auto-Related & 11 & -1.89 & 5 & 0.55 & 11 & -1.24 & 5 & 0.57 \\
			Utilities & 1 & 1.26 & 1 & 0.84 & 1 & 1.06 & 1 & 0.84 \\
			Oil \& Gas \& Pipeline & 6 & 0.65 & 8 & 0.00 & 7 & 0.14 & 8 & 0.00 \\
		\end{longtable}
		\par\footnotesize Distances are score minus the eighth-ranked scenario cutoff, in basis points of expected loss score.
	\end{landscape}
	
	\section{Additional sensitivity tables}\label{app:sensitivity}
	
	\begin{table}[H]
		\centering
		\caption{Random convex crosswalk-weight sensitivity}
		\label{tab:app_crosswalk_sens}
		\small
		\begin{tabular}{@{}lrrrrr@{}}
			\toprule
			Specification & 5th pct. & Median & 95th pct. & Minimum & Positive draws \\
			\midrule
			Expected loss & 2.7\% & 28.9\% & 34.4\% & approximately 0\% & 99.95\% \\
			Credit compensation & 62.9\% & 69.6\% & 72.2\% & 26.1\% & 100.0\% \\
			\bottomrule
		\end{tabular}
		\begin{minipage}{0.94\textwidth}\footnotesize
			\textit{Notes:} 2,000 independent Dirichlet(1,...,1) weight draws are generated within each multi-sector SNC--NGFS mapping. A draw is held fixed across all four scenarios. Percentages are the reduction in maximum regret relative to symmetric scenario weighting.
		\end{minipage}
	\end{table}
	
	\begin{table}[H]
		\centering
		\caption{Expected-loss sensitivity to Merton inputs}
		\label{tab:app_merton_sens}
		\small
		\begin{tabular}{@{}lrrrr@{}}
			\toprule
			Perturbation & Weighted baseline PD & Average max regret & MMR max regret & Reduction \\
			\midrule
			Risk-free rate 2.0\% & 0.0609\% & 0.100 & 0.072 & 28.6\% \\
			Risk-free rate 3.47\% & 0.0592\% & 0.097 & 0.069 & 29.1\% \\
			Risk-free rate 5.0\% & 0.0574\% & 0.093 & 0.066 & 29.6\% \\
			Debt-to-capital -10\% & 0.0425\% & 0.060 & 0.042 & 29.6\% \\
			Debt-to-capital +10\% & 0.0789\% & 0.144 & 0.103 & 28.5\% \\
			Equity volatility -10\% & 0.0220\% & 0.027 & 0.016 & 39.3\% \\
			Equity volatility +10\% & 0.1343\% & 0.297 & 0.227 & 23.5\% \\
			\bottomrule
		\end{tabular}
	\end{table}
	
	\begin{table}[H]
		\centering
		\caption{LGD sensitivity under both score specifications}
		\label{tab:app_lgd_sens}
		\small
		\begin{tabular}{@{}lrrrr@{}}
			\toprule
			Specification & LGD & Average max regret & MMR max regret & Reduction \\
			\midrule
			Credit compensation & 20\% & 0.000 & 0.000 & -- \\
			Expected loss & 20\% & 0.048 & 0.034 & 29.1\% \\
			Credit compensation & 40\% & 0.934 & 0.289 & 69.1\% \\
			Expected loss & 40\% & 0.097 & 0.069 & 29.1\% \\
			Credit compensation & 60\% & 0.000 & 0.000 & -- \\
			Expected loss & 60\% & 0.145 & 0.103 & 29.1\% \\
			\bottomrule
		\end{tabular}
		\begin{minipage}{0.94\textwidth}\footnotesize
			\textit{Notes:} Under expected loss, a common LGD scales all scores and therefore leaves the percentage regret reduction unchanged. Under the credit-compensation score, changing LGD can change sector rankings; ``--'' denotes a case in which scenario-conditioned allocations coincide to numerical precision and a relative reduction is not economically defined.
		\end{minipage}
	\end{table}

\end{document}